# Passively Q-switching cylindrical vector beam fiber laser operating in high-order mode

Hongxun Li,[1,2,3,4,†] Ke Yan,[1,†] Yimin Zhang,[1] Zhipeng Dong,[1] Chun Gu,[1] Peijun Yao,[1] Lixin Xu,[1,*] Rui Zhang,[2] Jingqin Su,[2,3,4,*] Wei Chen,[5] Yonggang Zhu,[5] and Qiwen Zhan[6]

[1]Department of Optics and Optical Engineering, University of Science and Technology of China, Hefei 230026, China
[2]Research Center of Laser Fusion, China Academy of Engineering Physics, Mianyang 621900, China
[3]Science and Technology on Plasma Physics Laboratory, Mianyang 621900, China
[4]Collaborative Innovation Center of IFSA (CICIFSA), Shanghai Jiaotong University, Shanghai 200240, China
[5]Jiangsu Hengtong Optical Fiber Technology Com.Ltd, Jiangsu 215000, China
[6]Electro-Optics Program, University of Dayton, Dayton, Ohio 45469, USA
[†]These two authors contributed equally to this work.
*Corresponding author: xulixin@ustc.edu.cn; sujingqin@caep.ac.cn

**We experimentally demonstrate a linear-cavity all-fiber passively Q-switching cylindrical vector beam (CVB) laser operating in high-order mode. This CVB fiber laser operates without any mode converter which always leads to high insertion loss, and it can realize high efficiency. In this fiber laser, the stable Q-switching pulse is achieved with a slope efficiency of 39%. By properly adjusting the polarization controllers, radially polarized and azimuthally polarized beams can be obtained. Our work proves the feasibility of achieving the stable Q-switching CVB pulse with high-order mode directly oscillating, and it may have an enormous potential for enhancing the efficiency.**

Recent years have seen increased interest in cylindrical vector beams (CVBs) which have doughnut intensity profiles and axisymmetric distribution of polarization [1]. Under the high NA focusing, the CVBs with radial or azimuthal polarization can be tightly focused. Thus the CVBs have a vast range of application in optical tweezers [2], high resolution metrology [3], surface plasmon excitation [4], beam shaping [5], electron acceleration [6] and material processing [7]. A lot of continuous-wave (CW) CVB lasers have been demonstrated by employing spatial polarization selective elements, such as spatially-variable retardation plate [8], calcite crystals [9], sub-wavelength grating [10], spatial light modulators [11], etc. Compared with the solid-state laser, all-fiber lasers with the few-mode fiber Bragg grating (FMFBG) [12], long-period fiber grating (LPG) [13-15] and mode selective coupler (MSC) [16-17] have the superiority of high compactness and flexibility.

Many applications of CVBs, especially material processing need high energy pulse, thus the all-fiber Q-switching and mode-locked CVB lasers have attracted immense research interest. In previous work of our group, a variety of all-fiber pulse CVB lasers have been reported. An all-fiber Q-switching CVB laser was demonstrated by employing the tungsten disulphide ($WS_2$) as a saturable absorber (SA) [18]. An all-fiber actively mode-locked CVB laser was demonstrated based on a Mach–Zehnder modulator [19]. Subsequently, K. Yan et al. demonstrated the passively Q-switching fiber lasers with CVB output based on $Bi_2Te_3$ and carbon nanotube (CNT) [20-21]. However, the previous CVB fiber lasers oscillated in the fundamental mode with the lateral offset splicing spot as mode converter, which led to a high insertion loss and hampered the enhancement of efficiency. Thus a high-order mode directly oscillated CVB fiber laser has recently been the subject of intensive research efforts, particularly for achieving CVB pulses. An all-fiber $LP_{11}$ mode oscillated CVB laser with high efficiency and low threshold was proposed by our group via using the polarization dependence of FMFBG. But such laser is difficultly employed in passively Q-switching laser due to the high polarization sensitivity. Liu et al. [22] presented an all-fiber laser which could stably oscillate at $LP_{11}$ mode vary with the intra-cavity polarization state.

In this letter, we demonstrate a linear-cavity all-fiber Q-switching CVB laser. To our knowledge, this is the first report on a Q-switching CVB fiber laser with high-order mode oscillation. Compared with other Q-switching CVB fiber laser, this fiber laser operates without any mode converter which always leads to high insertion loss, and it can realize high efficiency. The repetition rate of the stable Q-switching CVB pulses varies from 44.18 kHz to 58.16 kHz with a slope efficiency of 39%. In addition, the high purity Q-switching

CVBs, such as radially polarized and azimuthally polarized beams, are observed after an external FMFBG.

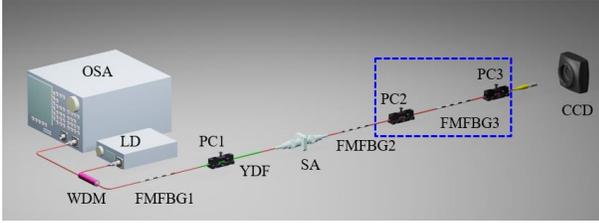

**Fig. 1.** Schematic illustration of the Q-switching CVB fiber laser.

Figure 1 shows the experimental setup of the linear-cavity all-fiber passively Q-switching CVB laser, which consists of a 974 nm laser diode (LD) pump, a 980/1060 nm wavelength division multiplexer (WDM), one section of Yb-doped fiber (YDF) of 30 cm length (core diameter of 10 μm and NA of 0.14), a fiber connector, a $WS_2$-PVA based SA similar to the reference [18], two FMFBGs with high reflectivity (FMFBG1, FMFBG3, fabricated in few-mode photosensitive fiber with core diameter of 9.07μm and NA of 0.117, reflectivity R=99.9%), an output coupling FMFBG (FMFBG2, fabricated in few-mode photosensitive fiber with core diameter of 9.61μm and NA of 0.108, reflectivity R=67%) and three polarization controllers (PC1, PC2 and PC3). The pump light from the LD is launched into the YDF via the WDM. The $WS_2$-PVA based SA is placed in the fiber connector which is located after the YDF to avoid amplification of the fundamental mode from mode coupling at the connector and improve the mode purity. FMFBG1 and FMFBG2 are used as the laser cavity mirrors and the mode selector for realizing high-order mode direct oscillation. PC1 in the laser cavity is employed to realize the stable Q-switching laser pulse output. In addition, a few external cavity elements are shown in blue dash frame, FMFBG3 is used as mode purifier to filter the fundamental mode component and obtain high purity CVBs, PC2 and PC3 are placed on each side of the FMFBG3 for adjusting the polarization of the output higher-order mode. All the fibers in the laser cavity are few-mode fibers which can support the $LP_{11}$ mode. The laser beam profile, output power, Q-switching pulse train and spectrum are monitored with a CCD camera, a powermeter, a 4 GHz digital oscilloscope (LeCroy, Waverunner 640Zi) and an optical spectrum analyzer (YOKOGAWA AQ6373B), respectively.

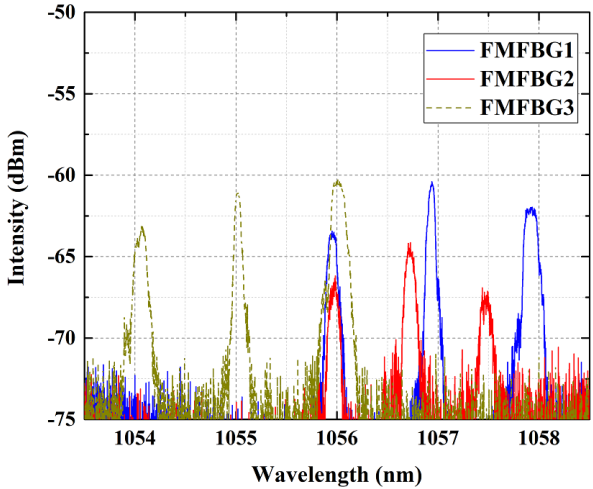

**Fig. 2.** The reflection spectra of the FMFBGs

The reflection spectra of FMFBG1 and FMFBG2 are shown in Fig. 2 (the blue and red solid lines), and the reflection spectrum of FMFBG3 is shown as the dark yellow dash line. The three reflection peaks in the reflection spectra of the FMFBGs indicate three kinds of reflection between different order modes. The right peak represents the $LP_{01}$ mode to the $LP_{01}$ mode reflection, the left peak represents the $LP_{11}$ mode to the $LP_{11}$ mode reflection, and the intermediate peak represents the $LP_{01}$ mode to the $LP_{11}$ mode reflection, respectively. FMFBG1 and FMFBG2 with the identical left peak are employed to realize $LP_{11}$ mode direct oscillation. And the right peak of FMFBG3 is coinciding with the left peaks of FMFBG1 and FMFBG2. Thus the fundamental mode component of the output is reflected, while the $LP_{11}$ mode component is exported.

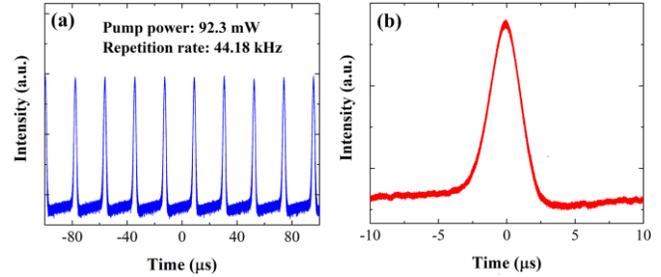

**Fig. 3.** (a) The Q-switching pulse train and (b) single pulse envelope under the pump power of 92.3 mW.

The threshold of the laser for continuous-wave (CW) operation is 57.0 mW. When the pump power reached 84.7 mW, the laser converts to a passively Q-switching mode. The repetition rates of the stable pulse train can be tuned by adjusting the pump power, which is a typical characteristic of the passively Q-switching laser. Stable Q-switching pulse is observed when the pump power gradually increases from 84.7 to 101.1 mW. Figure 3 shows the Q-switching pulse train and the single pulse shape of the Q-switching pulses with the pump power of 92.3 mW. The average output power is 13.24 mW with the pulse duration of 2.67 μs and the repetition rate of 44.18 kHz.

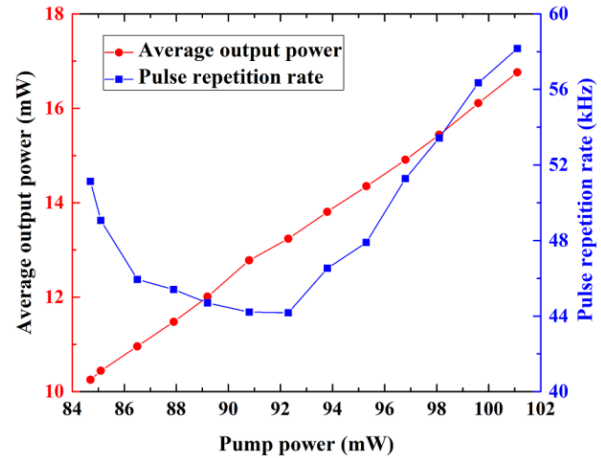

**Fig. 4.** The average output power and the pulse repetition rate of the Q-switching pulse versus the pump power.

Figure 4 shows the variation of the output power and the pulse repetition rate with the pump power. As the pump power increases from 84.7 to 101.1 mW, the output power increases from 10.25 mW to 16.76 mW, and efficiency of the Q-switching laser is as high as 39%. The pulse repetition rate descents at first and then increases, which coincides with the reference [18]. And the pulse repetition rate varies from 44.18 kHz to 58.16 kHz, with a tuning range 13.98 kHz. According to the relation between the average output power and the repetition rate, the maximum pulse energy is 299 nJ at the pump power of 92.3 mW.

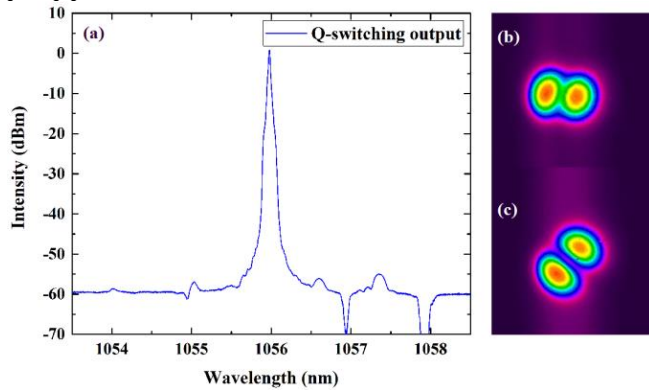

Fig. 5. (a) The optical spectrum of the laser from the pass port of the WDM; (b) the intensity profile of the output beam without the FMFBG3; (c) the intensity profile of the output beam after the FMFBG3.

The laser spectrum is measured from the pass port of the WDM. From Fig. 5(a) (blue solid line), the 30 dB linewidth of the Q-switching laser spectrum is less than 0.19 nm with the signal-to-background ratio of more than 55 dB. In addition, the oscillation wavelength is 1055.98 nm which exactly coinciding with the left peaks of the FMFBG1 and FMFBG2, indicating $LP_{11}$ mode directly oscillates in the laser cavity. However, the fiber connector results in the mode field mismatch which excites the $LP_{01}$ mode and reduces the $LP_{11}$ mode purity significantly. Thus the intensity distribution of the output beam directly from FMFBG2 is in the shape of peanut (shown in Fig. 5(b)). To improve the purity of the output mode, the external FMFBG3 is used as mode purifier for filtering the fundamental mode component. The oscillation wavelength is consistent with the right peak of the FMFBG3 so that the fundamental mode component from the mode field mismatch in the fiber connector can be well reflected. And high purity $LP_{11}$ mode output can be achieved after the external FMFBG3 and shown in Fig.5(c).

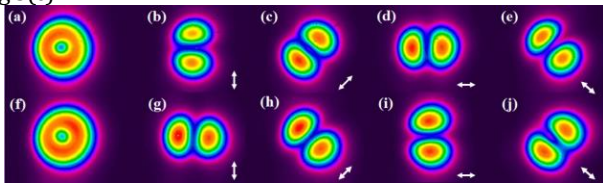

**Fig. 6.** (a) Beam profile of radial mode without a linear polarizer; (b)–(e) Transmitted intensity distributions of radial mode after a linear polarizer; (f) Beam profile of azimuthal mode without a linear polarizer; (g)–(j) Transmitted intensity distributions of azimuthal mode after a linear polarizer. The white arrows indicate the corresponding axis of the polarizer.

With properly adjusting PC1 and PC2, high purity CVBs (radially polarized and azimuthally polarized beams) can be obtained. And the beam profiles of the radially and azimuthally polarized modes are captured using the CCD camera as shown in Fig. 6(a) and (f), respectively. The doughnut intensity distribution is the typical feature of CVBs. And the output beams are confirmed to be radially polarized and azimuthally polarized by recording the CCD images with rotating a linear polarizer which inserted between the output port and the CCD camera (shown in Figs. 6(b-e) and 6(g-j)).

According to the results of the experiment, we can find that the SA insertion scheme using a fiber connector is not the most suitable choice for high-order mode directly oscillated Q-switching CVB fiber laser due to the mode field mismatch which leads to the output mode purity decline. Considering the purity of the output CVBs, next step for us is to seek a way to realize the pulse CVB output without influencing the oscillation mode field, such as synchronous pulse pump and evanescent field modulation.

In conclusion, we experimentally demonstrate a linear-cavity all-fiber passively Q-switching CVB laser based on the $WS_2$ saturable absorber and prove the feasibility of achieving the Q-switching CVB pulse with $LP_{11}$ mode directly oscillating. Compared with other Q-switching CVB fiber laser, this fiber laser operates without any mode converter which always leads to high insertion loss, thus our work may have an enormous potential for enhancing the efficiency. The laser oscillates at a single wavelength of 1055.98 nm with 30 dB linewidth less than 0.19 nm, and the slope efficiency is 39%. By adjusting PC1, the stable Q-switching pulse can be obtained with a repetition rate tuning range of 13.98 kHz (from 44.18 kHz to 58.16 kHz). When the pump power is 92.3 mW, the corresponding repetition rate is 44.18 kHz with the pulse duration of 2.67 μs, and the maximum pulse energy is 299 nJ. Furthermore, radially polarized and azimuthally polarized beams are obtained by tuning the polarization controllers.

**Funding.** National Natural Science Foundation of China (NSFC) (61675188, 61475145); Open Fund of Key Laboratory Pulse Power Laser Technology of China, (SKL2016KF03)